\documentclass[sigconf]{acmart}
\AtBeginDocument{%
  \providecommand\BibTeX{{%
    \normalfont B\kern-0.5em{\scshape i\kern-0.25em b}\kern-0.8em\TeX}}}

\copyrightyear{2022}
\acmYear{2022}
\setcopyright{acmcopyright}\acmConference[CHASE'22]{15th International Conference on Cooperative and Human Aspects of Software Engineering}{May 21--29, 2022}{Pittsburgh, PA, USA}
\acmBooktitle{15th International Conference on Cooperative and Human Aspects of Software Engineering (CHASE'22 ), May 21--29, 2022, Pittsburgh, PA, USA}
\acmPrice{15.00}
\acmDOI{10.1145/3528579.3529177}
\acmISBN{978-1-4503-9342-3/22/05}

\usepackage[most]{tcolorbox}
\usepackage{fontawesome}
\usepackage{xspace}
\usepackage{array}

\newcommand{\surveyquote}[1]{``\emph{#1}''\xspace}

\definecolor{boxcolor}{RGB}{238, 223, 204} %
\DeclareRobustCommand{\mybox}[2][gray!10]{%
\begin{tcolorbox}[   
        breakable,
        left=0pt,
        right=0pt,
        top=0pt,
        bottom=0pt,
        colback=#1,
        colframe=black,
        width=\dimexpr\columnwidth\relax, 
        enlarge left by=0mm,
        boxsep=5pt,
        outer arc=4pt,
        boxrule=.5mm
        ]
        #2
\end{tcolorbox}
}

\usepackage{tikz}
\usepackage{pgfplots}
\pgfplotsset{width=10cm,compat=1.9}

\newlength{\myboxheight}
\def\mybarchart#1{
\resizebox {#1} {\myboxheight} {%
\begin{tikzpicture}[]
\definecolor{clr1}{RGB}{99,99,99}
\definecolor{clr2}{RGB}{240,240,240}
\begin{axis}[
      axis background/.style={fill=white!10, draw=white!50},
      axis line style={draw=none},
      tick style={draw=none},
      ytick=\empty,
      xtick=\empty,
      ymin=0, ymax=1, 
      xmin=0, xmax=1]
\addplot [
      ybar interval=.5,
      fill=black,
      draw=none,
]
	coordinates {(1,1) (1,1)}; 
\addplot [
      ybar interval=.5,
      fill=black,
      draw=none,
]
	coordinates {(1,1) (0,1)}; 
\end{axis}%
\end{tikzpicture}%
}%
}

\usepackage{stackengine}

\begin{document}

\title[How Developers and Managers Define and Trade Productivity for Quality]{How Developers and Managers Define and Trade \\ Productivity for Quality}

\author{Margaret-Anne Storey}
\affiliation{%
  \institution{University of Victoria}
  \city{Victoria}
  \state{BC}
  \country{Canada}
}
\email{mstorey@uvic.ca}

\author{Brian Houck}
\affiliation{%
  \institution{Microsoft Corp.}
  \city{Redmond}
  \state{WA}
  \country{USA}
}
\email{brian.houck@microsoft.com}

\author{Thomas Zimmermann}
\affiliation{%
  \institution{Microsoft Research}
  \city{Redmond}
  \state{WA}
  \country{USA}
}
\email{tzimmer@microsoft.com}

\renewcommand{\shortauthors}{Storey, Houck and Zimmermann}

\begin{abstract}

  Background: 
  Developer productivity and software quality are different but related multi-dimensional lenses into the software engineering process. The terms are used liberally in industry settings, but there is a lack of consensus and awareness of what these terms mean in specific contexts and  which trade-offs should be considered. 
  
  Objective \& Method: Through an exploratory survey study with developers and managers at Microsoft, we investigated how these cohorts define productivity and quality, how aligned they are in their views, how aware they are of other views, and if and how they trade quality for productivity. 
 
  Results: We find developers and managers, as cohorts, are not well-aligned in their views of productivity---developers think more about work activities, while more managers consider performance or quality outcomes. We find developers and managers have more aligned views of what quality means, with the majority defining quality in terms of robustness, while the timely delivery of evolvable features that delight users are also key quality aspects. Over half of the developers and managers we surveyed make productivity and quality trade-offs but with good reasons for doing so.  
  
  Conclusion:  Alignment on how developers and managers define productivity and quality is essential if they are to design effective improvement interventions and meaningful metrics to measure productivity and quality improvements. Our research provides a frame for developers and managers to align their views and to make informed decisions on productivity and quality trade-offs.

\end{abstract}



\maketitle

\section{Introduction}

Software organizations strive to develop software quickly and efficiently, while achieving and maintaining high levels of quality.  At several large companies we have worked with, we heard many discussions about how to improve ``productivity'' and ``quality,'' but these discussions, many of which aimed to propose actionable productivity and quality metrics, often lacked a clear definition of what these nuanced terms mean. 

Sparked in part by the needs of industry, developer productivity has been studied extensively in software engineering research in recent years. Researchers have reported on how developers perceive productivity, sharing that developers consider productivity not just about writing code or delivering features quickly, but is also related to their ability to deliver high-quality work~\cite{sadowski_software_2019}, their ability to work without interruptions~\cite{meyer_developers_2019}, and is impacted by their satisfaction and well-being~\cite{storey_towards_2019}, their happiness at work~\cite{graziotin2013}, and many other social and technical factors~\cite{murphyhill2019predicts, Wagner2018}. Despite these many research findings, how to define and measure productivity in a way that provides meaningful insights and pragmatically supports change remains challenging.

There are established formal models~\cite{nistala_software_2019} and definitions of software quality~\cite{kitchenham_software_1996}, and many proposed metrics ~\cite{pradhan_quality_2019} for measuring system and process quality, including defects in delivered code, unplanned rework, code churn, reusability, code readability, test coverage, how code is reviewed, and customer satisfaction. It is also an ongoing debate in practitioner blogs which metrics to use\footnote{\url{https://www.bmc.com/blogs/software-quality-metrics/}, \\ \url{https://xbosoft.com/software-qa-consulting-services/definition-software-quality/}, \\ \url{https://www.testbytes.net/blog/what-is-software-quality/}}. Despite so many different views, Kitchenham and Pfleeger note that although it is not easy to define quality, we can recognize it when we see it~\cite{kitchenham_software_1996}. In our experience working in and with industry, how software developer practitioners define and think about quality in their day-to-day work varies greatly. To some, quality is about satisfying the visible functional requirements for a product, while to others it is more important to consider quality in terms of non-functional requirements, such as performance and reliability of the code or process. These different priorities for quality can lead to misaligned values and goals during the design and evolution of software, and can create tension among developers and managers when they try to balance developer productivity in terms of shipping high-quality features or completing tasks. 

In this paper, we present the findings from a study at Microsoft that investigated how developers and managers define and trade off productivity and quality. Through a survey answered by 165 participants, we consider whether developer views of productivity and quality are aligned with each other and with the views of managers. We further explore how developers and managers think about quality, whether they report making trade-offs between them, and why those trade-offs are made. 

Our research has revealed that developers and managers, as cohorts, are not well-aligned in their views of what it means to be \textbf{productive}. Notably, more developers think of productivity in terms of \textbf{activities} and task completion, while more managers think of productivity in terms of \textbf{performance and quality of their outcomes}. We also found that developers are not accurate at predicting their managers’ views of what it means for their team to be productive. 

In terms of \textbf{quality}, we found that although individual developers and managers have quite varied views of what quality means, the two cohorts are closely aligned in these different views, with the majority in both groups defining quality in terms of software \textbf{robustness}. We also found that over half of the developers and managers that answered our survey consider quality as something that can be \textbf{traded} for higher productivity. Conversely, approx. one-third of the developers and managers that answered our survey consider quality as a necessary part of productivity. 
We conclude with the recommendation that productivity and quality should be considered as related but multi-dimensional lenses into the software engineering process and that there is a need for more discussion by 
developers and managers to align on what these terms mean in their contexts and to use this understanding to drive how they approach measuring and changing how they approach improving productivity and quality.

\section{Methodology}

To understand how developers and managers define and trade productivity for quality, we conducted a survey study with employees at Microsoft. %
Our research questions were as follows: 

\begin{description}
\setlength{\itemsep}{0pt}
\setlength{\parskip}{0pt}
    \item[\textbf{RQ1}] {How do developers and managers define \textbf{productivity}? How \textbf{aligned} are developers and managers in their definitions of productivity?}
   \item[\textbf{RQ2}] {Do developers and managers \textbf{understand how the other cohort defines} productivity?}
    \item[\textbf{RQ3}] {How do developers and managers define \textbf{quality}? How \textbf{aligned} are developers and managers in their definitions?}
    \item[\textbf{RQ4}] {Do developers and managers \textbf{trade} quality for productivity? What \textbf{reasons} do they give for trading or not trading quality for productivity?} 
\end{description}

\subsection{Survey Design}
\label{sec:survey-design}

To help us prepare our survey questions, we conducted an exploratory focus group with six developers and two interviews with managers. 
We developed the survey in an iterative manner, with a pilot helping refine the survey questions. The survey was distributed in Jan 2020. We invited 1,500 developers and 720 managers at Microsoft to participate in the survey via
personalized email. The survey was advertised as a general survey on software productivity. After completion of the survey, participants could enter a sweepstakes to win one of four \$100 Amazon.com Gift Certificates. The survey was open for two weeks, and we received 131 responses from developers and 34 responses from managers (response rates of 9\% and 5\%, respectively). 
The ethics for this study were reviewed and approved by the Microsoft Research Institutional Review Board (MSRIRB), which is an IRB federally registered with the United States Department of Health \& Human Services (Reference: MSRIRB \#586). 
The survey instruments are available as supplemental material~\cite{supplemental-materials-alignment}.
In our initial interviews, we found that developers and managers believed they understood how each defined productivity, but they did not have a good sense of how the other cohort defined quality---so we omitted questions about how the other cohorts defined quality in our surveys.

The questions in the \textbf{developer} survey that we analyzed for this research are as follows: 
\begin{itemize}
\item When thinking about your work, how do you define productivity? 
\item How do you think your manager defines productivity?
\item When thinking about your work, how do you define quality?
\item Have you ever had to make trade-offs between productivity and quality? If so, please explain.
\end{itemize}
The questions we asked in the \textbf{developer manager} survey are: 
\begin{itemize}
\item When thinking about your team, how do you define productivity?
\item How do you think the developers you manage define productivity?
\item When thinking about your team, how do you define quality?
\item Has your team had to make explicit trade-offs between productivity and quality? If so, please describe those tradeoffs.
\end{itemize}

\subsection{Data Analysis}

We coded the open-ended responses to the survey in an iterative and inductive manner, where two paper authors coded all responses, then compared all codes in agreement sessions with all three authors present to refine the codes and check the coding of all responses. After three cycles of coding, we arrived at five codes for the productivity questions and quality questions. For the trade-off question, we noted several themes to describe why some developers and managers said they did or did not make trade-offs. 

The five main codes that emerged from coding the productivity definition question are as follows: developer \textbf{satisfaction and well-being}, \textbf{performance or quality outcomes of developer’s work}, the \textbf{activities} developers achieve in a certain time frame, the work they do \textbf{collaborating} with others, and the ability of developers to work \textbf{efficiently or with fewer interruptions}. 
We noted that these five emergent codes from our inductive coding aligned with the five dimensions proposed by the SPACE developer productivity framework~\cite{space} (see Table~\ref{table:space-codes}).

\begin{table}[ht]
\caption{Developer productivity codes and definitions~\cite{space}.}
\vspace{-0.5\baselineskip}
\label{table:space-codes}
\begin{tabular}{l l} 
\hline
S: & Satisfaction with work and personal well-being \\ 
P: & Performance and quality of development outcomes\\
A: & Activity as the count of actions or outputs \\
C: & Collaboration and communication among people \\ 
E: & Efficiency and flow of work with minimal interruptions \\
\hline
\end{tabular}
\end{table}

Five key codes also emerged from inductive coding of the quality questions (see Table~\ref{table:truce-codes} for a description of these codes). 
When the productivity and quality questions were coded, we counted the responses so that we could compare how aligned developers and managers were across the two cohorts and how closely they were able to predict how the other defined productivity and quality.  Some responses were coded with multiple codes. We show the frequency of the coded responses in the results below. To check alignment, we used Fisher's exact test to check for statistically significant differences for research questions 1, 2 and 3. We used Fisher's exact test rather than CHI square test due to our small sample size. 

\begin{table}[t!]
\caption{Quality codes and definitions.}
\vspace{-0.5\baselineskip}
\label{table:truce-codes}
\begin{tabular}{l l} 
\hline
T: & Timely and predictable delivery of features \\ 
R: & Robustness of code (reliable, tested, secure, scales, etc.) \\
U: & Meets user, customer, stakeholder requirements/needs \\ 
C: & Readable, documented code to facilitate collaboration \\ 
E: & Evolvable design with tests to support future changes \\
\hline
\end{tabular}
\end{table}

\subsection{Limitations}

Our exploratory study is a first step in understanding how productivity and quality are defined and traded by both managers and developers.  Future studies should address the following limitations. 

\textbf{External validity:} Since our survey was conducted at a single company, our results may not generalize to other companies. Furthermore, our sample was small and our results may not generalize to other developers and managers.  The response rate for the manager cohort was low.

\textbf{Construct validity:} The survey's open-ended questions may have been ambiguous and there may be response bias.  The order of questions may have influenced how our respondents answered subsequent questions in the survey (e.g., the trade-off question).  Furthermore, we asked the developers and managers to think about their own work and their team's work, respectively, which may have influenced how they answered questions about productivity and quality. Our wording was intentionally vague as the productivity and quality terms are often not clarified when used in industry.  

\textbf{Internal validity:} How we coded the questions may not match how other researchers would have coded our data, and we do not claim that our findings are reproducible. We tried to offset potential biases in how we coded the data by having two of us code all of the data independently. We then discussed our codes in agreement sessions with all three researchers present, with the third researcher breaking any ties in the case of disagreements (this was rare), but some responses were ambiguous and in those cases we did not assign codes to those responses. 
As this was conducted in a company, we do not have permission to share all responses but we include more examples of our coding in Tables~\ref{table:space-examples} and ~\ref{table:truce-examples}.

\section{Comparing How Developers and Managers Define Productivity} 

Through our surveys, we asked developers and managers to share their definitions of productivity as well as how they thought their managers or team members,  respectively, might define the term. 
As mentioned, we inductively coded these responses, finding that the top-level codes that emerged aligned with the five themes captured by the SPACE framework for developer productivity (see Table~\ref{table:space-codes}).  Examples of coded responses for productivity are shown in Table~\ref{table:space-examples}. 

\begin{table*}[ht]
\caption{Examples of how the responses to the questions about productivity were inductively coded.}
\vspace{-0.5\baselineskip}
\label{table:space-examples}
\begin{tabular}{ | m{6em} | m{13cm}| m{1.8cm} | } 
\hline
\textbf{Code} & \textbf{Example Quotes of How Developers Define Productivity} & \textbf{Participant ID} \\
\hline
\hline
Satisfaction and well-being & \surveyquote{My productivity is working on the real tool development and learning new skills.} & D63 \\
\hline
Performance & \surveyquote{Delivery of projects in the long term.} & D92 \\
\hline
Activity & \surveyquote{Number of tasks completed / iteration.} & D11 \\
\hline
Collaboration & \surveyquote{Ability to brainstorm high-quality ideas and provide feedback for teammates ideas, produce high-quality code/code design quickly and provide valuable feedback to code/code design of teammates.} & D114 \\
\hline
Efficiency and flow & \surveyquote{Having the information and tools needed to do my work efficiently. Having a quiet environment with minimal distractions to focus and remain in the flow state. Having as few meetings as possible.} & D84 \\
\hline
\end{tabular}
\end{table*}

\subsection{How Developers and Managers Define Productivity}
To answer RQ1, we asked Individual Contributors [ICs]~\footnote{We use the terms ICs and developers interchangeably throughout the paper.} and managers for their definitions of productivity (the exact wording of the questions in the survey is listed in Section~\ref{sec:survey-design}).
We show how their responses compare in Table~\ref{table:productivity-definitions}, and discuss their responses below.
Note we refer to quotes from specific developers and managers using D\# and M\#. 

\begin{table}
\caption{How ICs and managers define productivity.  The differences with how they defined productivity in terms of performance (P) and activity (A) are statistically significant (*).}
\vspace{-0.5\baselineskip}
\label{table:productivity-definitions}
\begin{tabular}{llll}
     \toprule
     & \stackanchor{ICs \textbf{define}}{own productivity} & \stackanchor{Managers \textbf{define}}{team's productivity} \\
     \midrule
     S & \mybarchart{8pt} 8\% & \mybarchart{9pt} 9\% \\ 
     P & \mybarchart{35pt} 35\% & \mybarchart{67pt} 67\% & (*) \\ 
     A & \mybarchart{50pt} 50\% & \mybarchart{21pt} 21\% & (*) \\ 
     C & \mybarchart{24pt} 24\% & \mybarchart{33pt} 33\% \\ 
     E & \mybarchart{38pt} 38\% & \mybarchart{45pt} 45\% \\
     \bottomrule
\end{tabular}
\end{table}

\subsubsection{How developers define productivity}
We note from Table~\ref{table:productivity-definitions} that developers are more likely than managers to consider productivity in terms of \textbf{activity} (writing code, fixing bugs, issuing pull requests, and other tasks completed in a given period of time).  For example, D21 described their productivity this way:  \surveyquote{How many artifacts I produce, example: Pull Requests, check-ins, dev docs, emails, etc.} But many consider productivity in terms of \textbf{performance}, such as quality of the work done or features delivered to the customers, as D98 shared: \surveyquote{How efficiently I'm completing my work items for the given sprint while maintaining a high level of quality in my code.} Many ICs, however, define their productivity in terms of \textbf{efficiency} and their ability to achieve a state of flow~\cite{mikkonen2016flow} while working. As D27 mentioned: \surveyquote{Percentage of my time spent doing actual work versus time spent in meetings, time spent doing manual tasks I shouldn't be doing, or wandering around looking for someone with subject-matter expertise (basically work versus meta-work).}

Some ICs also mentioned the importance of \textbf{collaboration}, for example, D119 defined their productivity as follows: \surveyquote{Amount of useful `work' (feature implemented, customer issues resolved, colleagues helped) done compared to `distractions' (interruptions, planning meetings, context switching).} This response was also coded with P and E. Only a few mentioned developer \textbf{satisfaction} and well-being as aspects of productivity, for example, D134 mentioned how being engaged is important to them: \surveyquote{I define productivity as how well I felt engaged in the work I am doing, as well as the lack of feeling stopped or held back. I attest to this as I notice if I am truly engrossed within a task I am just naturally more productive ( through emotion) and have a strong desire to solve [the] problem given to me.} 

\subsubsection{How managers define their team's productivity} 
When we asked managers to define productivity when thinking about their team, their responses were quite different to how ICs define productivity (see Table~\ref{table:productivity-definitions}).  Managers are more likely to define productivity for their team in terms of \textbf{performance} and/or \textbf{efficiency} rather than \textbf{activity}.  M20 shared that their view of their team's productivity is to: \surveyquote{Tackle the right problem and get the job done efficiently \& high quality}. Note we coded this response with P and E. 

About one-third of the respondents to the manager survey focused on \textbf{collaboration} as an important aspect of productivity.  M18 mentioned: \surveyquote{Being able to get out of meetings with action items, and proper end result. Having the right folks in the room so that we can close on things and move on.} Some of the manager responses for their definition of their team productivity crosscut several dimensions, as M6 noted:  \surveyquote{People are able to predictably deliver features and fixes that keep our customers happy while learning and growing, constantly improving our culture, and staying happy themselves} (this response was coded with S, P, C, and E).

\mybox{\faArrowCircleRight~\textbf{Takeaway:} Developers and managers defined productivity in terms of all five dimensions of SPACE, but only a few mentioned developer satisfaction and well-being. Notably, developers and managers, as cohorts, were not that closely aligned in their views of productivity as \emph{developers are more likely to define productivity in terms of activity, while managers are more likely to define productivity in terms of performance.}}

\subsection{How Developers Think Managers Define Productivity}

To answer RQ2, we used the same five codes as RQ1 to code responses to the question we posed to developers about how they think their managers define productivity. Table~\ref{table:dev-view-mgr-productivity} shows that the IC perceptions of how managers define productivity does not align with how managers define productivity, especially with respect to the distributions of answers coded with Performance, Activity and Efficiency. Managers are more likely to think about productivity in terms of \textbf{performance} (as discussed above), but ICs anticipate managers define productivity more in terms of \textbf{activity}.   For example, D82 is one of many developers that considers that their manager defines productivity in terms of their \textbf{activity} only: \surveyquote{codes have been checked in, bugs have been fixed.} Some do consider that their managers define productivity in terms of \textbf{performance}, and a few anticipate that managers consider \textbf{collaboration} as an aspect of productivity.  For example, D81 responded: \surveyquote{How much impact is created. How much [value] added to others work and how much of others work is leveraged} (coded with both P and C). Furthermore, some developers underestimated the number of managers that consider \textbf{efficiency} as an important aspect of productivity. Only a few mentioned that their managers consider \textbf{satisfaction} and well-being, with one of the few notable exceptions by D44: \surveyquote{making sure all are happy.}  
Of note is that five ICs were not sure how their managers define productivity, suggesting that a conversation about productivity had not taken place.

\mybox{\faArrowCircleRight~\textbf{Takeaway:} Developer anticipation of how their managers define productivity is not aligned with how managers actually define productivity, as \emph{managers are more likely to define productivity in terms of performance rather than activity}.  \emph{Developers also underestimated how often managers mentioned efficiency} as an important aspect of productivity.}

\begin{table}
\caption{How ICs think managers define productivity compared with how managers actually define productivity. The differences in terms of performance (P), activity (A), and efficiency (E) are statistically significant (*).}
\vspace{-0.5\baselineskip}
\label{table:dev-view-mgr-productivity}
\begin{tabular}{llll}
     \toprule
     & \stackanchor{ICs \textbf{think} managers}{define productivity} & \stackanchor{Managers \textbf{define}}{team's productivity} \\
     \midrule
     S & \mybarchart{5pt} 5\% & \mybarchart{9pt} 9\% \\ 
     P & \mybarchart{37pt} 37\% & \mybarchart{67pt} 67\% & (*) \\ 
     A & \mybarchart{53pt} 53\% & \mybarchart{21pt} 21\% & (*) \\ 
     C & \mybarchart{19pt} 19\% & \mybarchart{33pt} 33\% \\ 
     E & \mybarchart{12pt} 12\% & \mybarchart{45pt} 45\% & (*) \\
     \bottomrule
\end{tabular}
\end{table}

\subsection{How Managers Think Developers Define Productivity}

From Table~\ref{table:mgr-view-dev-productivity} we can observe that managers, as a cohort, are quite accurate at estimating how ICs define productivity (RQ2).  This contrasts with how many ICs do not have such an accurate perception of how managers define productivity.  Many managers anticipate that many ICs will define productivity in terms of activity. For example, M37 suggested that ICs on their team define productivity in terms of \textbf{activity} in contrast to how they (as managers) or their peers define productivity: \surveyquote{Achieving current sprint deliverables.  We use two weeks sprint cycles with monthly integration cycles. Direct reports are more focused along these more immediate time lines that myself or my peers.}  Many managers, however, recognize that their ICs think about efficiency and flow when defining productivity, as M11 mentioned their ICs consider: \surveyquote{How many hours a day can they spend in uninterrupted coding activity.} Of note, and as discussed above, is that few ICs thought their managers considered \textbf{efficiency} and flow in their definitions of productivity (see Table~\ref{table:dev-view-mgr-productivity}).  

\mybox{\faArrowCircleRight~\textbf{Takeaway:} Many \emph{managers have accurate insights into how ICs define productivity} (notably in terms of activity, performance and efficiency), even though those views are not well-aligned with their own definitions of productivity. }

\begin{table}
\caption{How ICs define productivity compared with how managers think ICs define productivity.  No differences are statistically significant.}
\vspace{-0.5\baselineskip}
\label{table:mgr-view-dev-productivity}
\begin{tabular}{llll}
     \toprule
     & \stackanchor{ICs \textbf{define}}{own productivity} & \stackanchor{Managers \textbf{think} ICs}{define productivity} \\
     \midrule
     S & \mybarchart{8pt} 8\% & \mybarchart{15pt} 15\% \\ 
     P & \mybarchart{35pt} 35\% & \mybarchart{24pt} 24\% & \\ 
     A & \mybarchart{50pt} 50\% & \mybarchart{52pt} 52\% & \\ 
     C & \mybarchart{24pt} 24\% & \mybarchart{12pt} 12\% \\ 
     E & \mybarchart{38pt} 38\% & \mybarchart{42pt} 42\% \\
     \bottomrule
\end{tabular}
\end{table}

\section{Comparing How Developers and Managers Define Quality}

We asked developers and managers how they define quality (to answer RQ3). 
We did not ask them how they anticipated the other cohort may define quality as we found in the exploratory interviews that asking about their own views of quality was already challenging to answer. 

\subsection{How Developers and Managers Define Quality} 
\label{sec:quality-definitions}

Five key codes emerged from our inductive coding of the responses to the questions  ``when thinking about your work, how do you define quality'' posed in the developer survey, and ``when thinking about your team's work, how do you define quality'' posed in the manager survey.  The five core codes are shown in Table~\ref{table:truce-codes} above, and samples of coded responses are shown in Table~\ref{table:truce-examples}.

\begin{table*}[ht]
\caption{Examples of how responses to the question about quality were inductively coded.}
\vspace{-0.5\baselineskip}
\label{table:truce-examples}
\begin{tabular}{ | m{12.5em} | m{10cm}| m{2cm} | } 
\hline
\textbf{Code} & \textbf{Example Quotes of How Developers Define Quality} & \textbf{Participant ID} \\

\hline
\hline
Timeliness of features delivered & \surveyquote{Shipping code that works, and on time.} & D13 \\
\hline
Robustness of features & \surveyquote{Robustness of code.} & D9 \\
\hline
User delight & \surveyquote{Degree to which the product meets customer needs.} & D46 \\
\hline
Collaboration needs met & \surveyquote{Quality is code or solutions that solve a problem and don't need undue maintenance or lengthy handoff. If I get hit by a bus and the company can still easily use the code I've written, I've made a quality solution.} & D129 \\
\hline
Evolution needs met & \surveyquote{Easy to test, easy to change.} & D15 \\
\hline
\end{tabular}
\end{table*}

\begin{table}
\caption{How individual contributors (ICs) and managers define quality. The differences with how they define quality in terms of robustness (R) are statistically significant (*).}
\vspace{-0.5\baselineskip}
\label{table:quality-definitions}
\begin{tabular}{llll}
     \toprule
     & ICs \textbf{define} quality & Managers \textbf{define} quality &  \\
     \midrule
     T & \mybarchart{7pt}   7\% & \mybarchart{15pt} 15\% & \\ 
     R & \mybarchart{71pt} 71\% & \mybarchart{88pt} 88\% & (*) \\ 
     U & \mybarchart{38pt} 38\% & \mybarchart{39pt} 39\% & \\ 
     C & \mybarchart{17pt} 17\% & \mybarchart{18pt} 18\% & \\ 
     E & \mybarchart{44pt} 44\% & \mybarchart{33pt} 33\% & \\
     \bottomrule
\end{tabular}

\end{table}

The most common theme (see Table~\ref{table:quality-definitions} for the distribution of answers coded with the five emergent quality codes) for both developers and managers is defining quality in terms of the \textbf{robustness} of the software, which can refer to reliability, performance, how well tested the code is, and adherence to security concerns. 

The next most frequent response for both developers and managers referred to how \textbf{evolvable} the software is (i.e., how its design and test suites support future changes).  For example, D42 felt that quality meant \surveyquote{maintainable, and defect free}, while M23 mentioned that \surveyquote{quality is work that is designed for the future, unit and manually tested} (both responses were coded with Robust (R) and Evolvable (E)).  A similar relative number of developers and managers defined quality in terms of the delivery of features that \textbf{meet user needs} (coded with U). As M19 put it: \surveyquote{Quality is doing what the customer wants.} 

Some developers and managers noted that quality is about writing software that others, not the just authors, can build on explicitly and thus support \textbf{collaboration}. As D112 put it: \surveyquote{well documented code that is easy for newcomers to the team to get up to speed with.} And lastly, a few developers mentioned the \textbf{timeliness} of the features that are delivered as a core aspect of quality, as M26 said: \surveyquote{less bugs after release, followed by time to market} (coded with R and T).  

Some of the open-ended responses we received to this question addressed three or more of the five dimensions of quality that emerged from our coding. For example, M11’s response to this question was coded with R, U and E:  \surveyquote{Quality isn't the absence or pretense of some metric, rather it is an emergent property of a complex system. The property incorporates some sense of `bug free' and `suitability for a specific purpose' and also includes a bunch of other things like maintainability, diagnosability, security, robustness, availability, etc.} And D84's definition captures all five codes (T, R, U, C, E): \surveyquote{Available to customers at the originally planned date. Design that is clear to peers who will review the final work (code). Bug free and with full test coverage. Well documented for other peers to understand. Meets all planned and emergent requirements.}

\subsection{How Aligned Are Developers and Managers in How They Define Quality}

We compared how the two cohorts were aligned in their views of quality. Interestingly, and surprising to us, developers and managers as cohorts considered quality in similar ways in terms of distribution of the different responses, but within each cohort there were varied views of what quality refers to (as presented above, some definitions of quality were quite narrow, while others covered multiple dimensions). We expected and did see more quality definitions concerning Robustness (R), Evolvability (E), and meets User needs (U). Fewer managers and developers mentioned Timeliness (T) and Collaboration(C), but those that did clearly articulated how these aspects were very relevant to quality according to their view. Although more managers mention robustness than developers (proportionally), managers were also more likely to provide more elaborate definitions of quality.

\mybox{\faArrowCircleRight~\textbf{Takeaway:} Developers and managers defined quality, as cohorts, with the same inductively derived five themes (timeliness, robustness, delights users, meets collaboration needs, supports evolution). Over 70\% of both cohorts mentioned robustness in their definitions of quality, while around a third mentioned ``delights users'' and is ``evolvable''. }

\section{Do Developers and Managers Trade Quality for Productivity?}

We asked developers and managers, through an open-ended question, if they or their teams traded quality for productivity (per RQ4). We classified their answers into ``Yes, they report trading quality in favour of productivity'', or ``No'' --- they either report not trading quality for productivity, or they are more likely to trade productivity (in terms of feature delivery) for higher quality.  For the two cohorts, we saw a similar distribution of responses to this question (see Table~\ref{table:tradeoff-numbers}).  We further analyzed their open-ended responses to identify themes to explain their answers. We share insights from this analysis below.

\subsection{Why and How Developers and Managers Trade Quality for Productivity}
Over half of the developers and managers indicated they trade quality for productivity. 

Many reported that \textbf{meeting deadlines led to lower quality}.  D36 mentioned there is \textbf{insufficient time prioritized for quality} goals: \surveyquote{All the time. High-quality work takes much longer during all phases of development (design, implementation, testing, documentation) and there is seldom sufficient time available to meet productivity goals if quality were a priority.}  D13 also mentioned they had insufficient focus time to meet their deadlines: \surveyquote{Yes. I often have to take shortcuts to meet tight deadlines. The constant randomization my team faces is the main driver of this.}  Note that D13 recognizes that interruptions may lead to quality trade-offs. 
In contrast, D7 clarified that \textbf{meeting deadlines is itself a form of quality}: \surveyquote{
SW engineers including me have a tendency to over complicate solutions and these short-solutions keep us grounded on the reality that code that runs \_right\_now\_ has a quality element of its own!} 

Many participants also explained that working with \textbf{legacy code is a barrier to improving quality}, as D11 shared: \surveyquote{Yes all the time. The code base is so old and there're lots of inefficient / misleading implementation. It's almost impossible to fix them at once and no one dares to do so. So we're always sacrificing quality for productivity.}  Furthermore, some mentioned a need to incur \textbf{new technical debt} to meet deadlines, as D86 shared: \surveyquote{Yes, occasionally it becomes necessary to take on technical debt in order to ship a feature. This is ok as long as it leads to a better long-term return […].} However, D21 noted that this trade-off can backfire: \surveyquote{That's technical debt. Our product is a monument to what several decades of trading productivity for quality look like. Most of the time, it's impossible to predict how quality will suffer as a result of productivity and so retrospectives take several years to fully develop. But yes, I'll often favor finishing work over doing it `right' the first time because it needs to go out the door.}  Several respondents described that adding new technical debt was the result of \textbf{reduced or deferred testing} with potential long-term impacts on the evolvability of the code, as D44 shared: \surveyquote{Often, for example when building test infrastructure will take more effort than feature itself, we make a trade-off towards simpler test coverage, that hits us in the head later.}

\begin{table}[t!]
    \centering
    \caption{Responses to the question ``Do you trade quality for productivity?''  }
    \vspace{-0.5\baselineskip}
    \label{table:tradeoff-numbers}
    \begin{tabular}{@{}llll@{}}
    \toprule
         & Yes & No & Unknown \\
         \midrule
         Developers & \mybarchart{52pt} 52\% & \mybarchart{34pt} 34\% & \mybarchart{14pt} 14\% \\
         Managers & \mybarchart{59pt} 59\% & \mybarchart{38pt} 38\% & \mybarchart{3pt} 3\% \\
    \bottomrule
    \end{tabular}
\end{table}

Incurring technical debt and deferring testing were seen as necessary for meeting short-term needs such as \textbf{unblocking others}, as M21 shared: \surveyquote{Bootstrapping a new product on which hundreds of developers are dependent, we will bring up an MVP that largely works, but has non-mainline blocking bugs so they can start moving sooner rather than later}; or to quickly \textbf{meet user needs} and speed up feedback, as D117 said: \surveyquote{sometimes we are creating feature that are `scenario enabler’, then we can iterate over to make it usable with the best customer experience.} M29 shared there are processes in place to ensure \textbf{short-term quality trade-offs are not permanent}: \surveyquote{We work in an environment where we are encouraged to `run with safety scissors' - we're encouraged to take risks but do so in a safe way, where rollbacks, failovers, or recovery is possible and quick. So I wouldn't say we make trade-offs on quality, but if we do it is usually short-term with a plan to improve and seek feedback along the way.}  

\textbf{Culture around quality} also played a role in priorities of feature delivery over quality, as D54 shared: \surveyquote{Yes. Again this is a theme on team culture. Quality was secondary. The top priority was getting the feature to work.} In contrast, D102 said that although there are pressures to make compromises, trade-offs are mostly imagined: \surveyquote{There's always perceived pressure to make compromises on quality in favor of speed, but I think it's mostly imagined.}   Several participants also noted a tension in terms of the \textbf{return on investment} of spending too much time on quality over faster feature delivery to their customers, as D103 shared: \surveyquote{Always. It's the primary rub of my craft. The business needs to reach the customer quickly. There are only so many  hours in the day to do everything required for perfect quality. Typically there's a line where `good enough' is good enough and preferable to quality perfection. At some point, further investments in quality have diminishing returns for customer experience.} 

\begin{figure*}[ht]
\centering
\vspace{-\baselineskip}
\includegraphics[width=22cm]{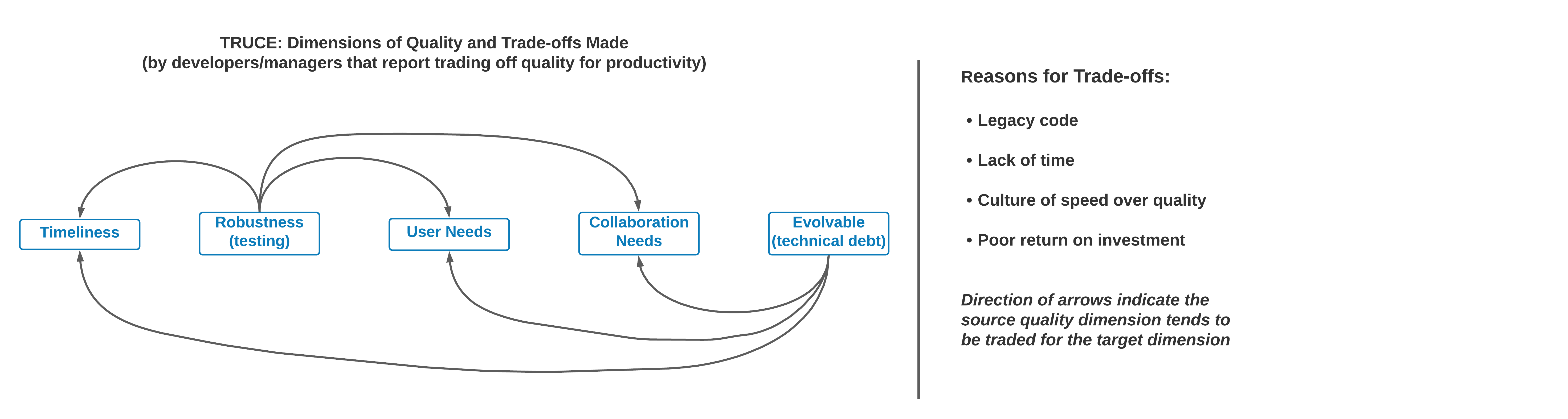}

\vspace{-\baselineskip}
\caption{Trade-offs that developers and managers reported making when asked about trading off quality for productivity.  We note that many reported quality trade-offs rather than quality for productivity. }
\label{figure:tradeoff-graph}
\end{figure*}

Although when we asked about how productivity is traded for quality, we found that the trade-offs shared were more about different dimensions of quality rather than productivity (see Fig.~\ref{figure:tradeoff-graph}). 

\mybox{\faArrowCircleRight~\textbf{Takeaway:} Over 50\% of developers and managers reported \emph{trading quality for productivity}, most to meet \emph{short-term needs}. \emph{Technical debt} and \emph{insufficient testing} were reported as trade-offs to achieve higher productivity with \emph{timely feature delivery} that \emph{meets user needs} and \emph{unblocks others}. \emph{Legacy code, lack of time, poor return on investment and a culture} that prioritizes faster feature delivery over quality were drivers for these trade-offs.}

\subsection{Why Some Developers and Managers Do \emph{Not} Trade Quality for Productivity}

Notably, approximately one-third of both developers and managers said that they \textbf{do not trade quality for productivity and rather reduce feature delivery to achieve quality}.  Bluntly, D52 shared: \surveyquote{The question is invalid.  Delivering lower quality work in order to deliver a larger number of features is not productivity} and M17 shares: \surveyquote{No, we do not cut edges. We rather be later than wrong.}   

Some specifically declared that they consider \textbf{quality as a part of productivity}. D126 said: \surveyquote{No.  For my job, improving quality IS productivity}; and M6 described: \surveyquote{Sort of. If you just think of productivity as `how many features did I ship' then yes, it's normal to trade off features vs quality, usually by spending more time on quality and shipping fewer features and unfortunately sometimes the other way around. But I would normally think of the activities that create quality (bug fixing, testing, better design) as just another part of productivity--if someone is getting those things done, I see them as being very productive. So In that sense, no, there's no trade-off, because quality is part of productivity.}

M31 mentioned aiming for \textbf{quality over feature delivery with a new product}: \surveyquote{Absolutely. We're shipping v1 of our product soon, and we're terminating feature work three months in advance to make sure that we can truly ship a quality product.} D15 mentioned that \textbf{even with legacy code, quality is important}: \surveyquote{legacy code that is known to be obsolete but must be fixed} (note legacy code was in contrast sometimes noted by others as a reason to trade feature delivery over quality above). 

Some respondents mentioned that \textbf{trading quality with feature delivery can impact developer satisfaction}. We saw two camps in these responses.  For some, they indicated that if they have to trade quality for productivity, they would be less satisfied with their work.  As D62 noted:  \surveyquote{No, [trading quality for productivity is] the reason for bad work life balance.}   D33 felt so strongly about this that they shared they are willing to spend more time to ensure they also remain productive: \surveyquote{I prioritize quality even if I may have to take extra time to ensure that my code is of good quality.}
Others, however, find prioritizing quality over productivity is less rewarding, as D53 shared: \surveyquote{I have been in this team for 4 months now and I find it very hard to find myself productive because my new team wants things a certain way and they are caring a lot about the code I put in. Their definition of quality is definitely affecting my productivity and hence my motivation.}  
And D79 reported they observed frustration with others: \surveyquote{Yes, in past projects I have had to implement some form of ask mode to ensure that we had enough time to stabilize the code we had already written.  This often frustrated engineers who did not want to have their checkins gated on lead/manager approval.}  That is, not making progress on features in favour of higher quality lowered these developers’ satisfaction. 

In our analysis, we observed some \textbf{tensions with manager priorities} for balancing quality and productivity.  D95 shared: \surveyquote{I'm sure my managers at times would've liked me to sacrifice quality to just get a feature done, but I honestly don't really know how to do that. I usually use tests to ensure that the code that I write does what I want it to do, and generally that `once it's done, it's done' feeling comes from writing good test cases.}  D129 shared how they also push back: \surveyquote{There's often a push for getting things out quicker than is possible which often means taking shortcuts and not documenting fully. I try to push against this and make solutions that follow best code and documentation practices.} In contrast, D113 shared: \surveyquote{My manager consistently demonstrates the willingness to trade productivity for quality; typically by avoiding being date-driven.}  

\textbf{Trade-offs between quality and productivity may need to be considered across teams.} M5 shared: \surveyquote{If i feel that our quality isn't where it needs to be, we will swarm on whichever part of the system needs it. That means productivity for those other areas will slow down, temporarily. I've worked on projects where the priority has been to help another team's quality and we've effectively stopped our work to help them because the release needed it. Another example might be ramping other resources for better coverage in our codebase: the SME's personal productivity will slow but the benefit is that we'll have multiple people now aware of that area.}  

\mybox{\faArrowCircleRight~\textbf{Takeaway:} Approximately one-third of developers and managers indicated they \emph{do not trade quality for productivity}, as they consider \emph{quality an essential aspect of productivity (for new and legacy systems)}. However, \emph{trading feature delivery for more quality can positively or negatively impact developer satisfaction}, depending on \emph{team and manager priorities}.}

\section{Discussion}

We discuss how the main dimensions in the SPACE framework mapped to the definitions developers and managers provided to us in the survey, how the responses to the quality question led to a new framework of thinking about software quality that we call TRUCE, and how SPACE and TRUCE are related but different lenses for thinking about software engineering processes.

\subsection{SPACE: Comparing How Developers and Managers Define Developer Productivity}
We inductively coded the responses about productivity to see how developers and managers in a company define productivity and if they are aligned in their views. 
 We noticed in our analysis that the five main codes that emerged mapped 1-1 to the five dimensions proposed by the SPACE framework~\cite{space}, a framework we co-authored. The SPACE framework emerged from the combined insights across several studies rather than from a single empirical study.  However, the SPACE framework was developed some months after we conducted this study and after we initially analyzed our data. Later we discovered that the five categories that emerged from our study data aligned with the five dimensions in SPACE.
Our study adds support for the dimensions of productivity captured in the SPACE framework and that the SPACE framework adequately captures how developers and managers, collectively, define productivity.

However, as we can see from Table \ref{table:dev-view-mgr-productivity}, the majority of the developer definitions of productivity concerned \textbf{activities}, while the majority of manager definitions mentioned \textbf{performance} or quality outcomes. And across both cohorts, fewer developers and managers considered other dimensions such as developer satisfaction, collaboration and efficiency. A few, however, mentioned 4-5 of the five dimensions.
This misalignment within and between cohorts shows that more discussion about what productivity means is needed, especially if productivity improvement goals and metrics are being proposed. There is also a need for awareness of how others define productivity (see Table \ref{table:mgr-view-dev-productivity}) to avoid potential false assumptions and later conflicts about how to improve and measure productivity.

\subsection{TRUCE: A Framework for Software Quality}
We were interested to learn how developers and managers define quality in their own words. From our inductive coding of the responses we received to the quality question, five main codes emerged (after several iterative rounds of coding and comparison across coders). Although it was not our intent when we designed this study to derive a new definition for quality, we offer TRUCE as a  framework for thinking about quality (similar to the SPACE framework for productivity). However, since it emerged from a single study, it would need to be validated in future studies. We note that the dimensions in this framework come from the “mouths” of the developers and managers we surveyed. TRUCE captures five main quality dimensions: 
The \textbf{Timely} delivery of 
\textbf{Robust} features that 
meets \textbf{User needs},  
while enhancing \textbf{Collaboration} with others 
and supporting the product’s future \textbf{Evolution}.

As shown in Table~\ref{table:quality-definitions}, fewer responses to the quality definition question mentioned C (collaboration) and T (timeliness), and yet these aspects are critical aspects of quality that should be considered and surfaced more often. Enabling collaboration or use by others is often invisible and hard to measure; this may explain why it was rarely mentioned, and yet it is very important in many development contexts.  Similarly, timeliness was infrequently mentioned; we were surprised to see this code surface in our analysis. 
Yet the responses convinced us that timeliness is indeed a very important aspect of quality. 
Similar to the breadth of productivity definitions, there is misalignment within cohorts of how developers and managers define quality (Table \ref{table:quality-definitions}).  The majority mention \textbf{robustness}, but the other dimensions also come into play when others think about what quality of the product and their process means to them. 

\subsection{SPACE and TRUCE: Related but Different Lenses Into Software Development}

Productivity is a lens that focuses on the developer, while quality is a lens that focuses on the quality of the software produced and process used. Although different lenses, they are related. In fact, when we asked about trade-offs between quality and productivity, we found that the trade-offs mentioned were more about trading quality aspects (see Fig.~\ref{figure:tradeoff-graph}) than trading quality for productivity. 

This is not surprising as there is overlap across the dimensions in SPACE and TRUCE.  Collaboration is in both frameworks: in SPACE, it captures the idea that one can’t think about productivity without considering how what they do is part of a bigger collaborative effort (e.g., code reviews), while in TRUCE, it captures how the quality of the software facilitates other's work (e.g., documented code that can be extended by others). Furthermore, the Performance dimension in SPACE captures the quality of the work that is done.   
Although there is some overlap, the two frameworks can be used by developers and managers to reflect on and discuss trade-offs between productivity and quality, and potentially to define metrics (see ~\cite{space} for a discussion of productivity metrics). 
As M37 put it:  \surveyquote{We are often/always constrained on the challenge of quality vs. time vs. cost/resources.  [...]  Productivity is a constant and elusive trade-off between immediate and future throughput.}  

\section{Related Work}

Before concluding our paper, we map our findings to related research on developer productivity and software quality. 

\subsection{Developer Productivity}

There has been extensive research in recent years on developer productivity and how system engineering activity metrics can provide important signals about developer activity and productivity.
Wagner and Ruhe's review of the literature summarize studies that use performance measures such as \textit{lines of code} or \textit{function points} as proxies to productivity~\cite{Wagner2018}.
However, many researchers and practitioners emphasize that developer productivity cannot and should not be measured by engineering metrics alone---development work is not mechanized work that can be assessed using a single metric, and doing so may be detrimental to overall and long-term development objectives~\cite{ko2019}.
For example, developers spend time mentoring newcomers, reviewing each other's work informally, and learning new skills. 
Most related to our work, Meyer et al.~\cite{MFMZ14} investigated how developers perceive and think about their own productivity. Meyer et al. also note the importance for developers to work without interruptions~\cite{meyer_developers_2019}. 

Murphy-Hill et al.~\cite{murphyhill2019predicts} asked 622 developers in a survey across three companies (Google, ABB and National Instruments) about productivity factors and self-rated productivity. They found that non-technical factors, such as job enthusiasm, peer support for new ideas, and useful feedback about job performance, correlated most strongly with self-rated productivity.
The three factors with lowest variance (across the three companies) were social and environmental rather than technical factors. Developer productivity and satisfaction has been discussed in conjunction with other human aspects of software engineering, such as developer happiness~\cite{graziotin2013} and developer motivation~\cite{BBHRS08, SBBHR09, FDS18}, while the impact of the pandemic on team productivity was discussed in~\cite{miller:icse:2021}. 
Some recent research led to the SPACE framework of productivity~\cite{space}---we saw that its five dimensions of productivity mapped to the different dimensions of productivity mentioned by our survey respondents.

\subsection{Software Quality}

Research about software quality has been extensive over the past few decades (and especially in the 1990s). But what quality means in any domain is not absolute---it remains somewhat elusive, despite the ISO standards for defining quality\footnote{See \url{https://asq.org/quality-resources/learn-about-standards}}. 
Some of the earliest research on software quality was done by Basili et al.~\cite{basili} and led to the powerful Goal-Question-Metric which still stands the test of time in terms of defining quality metrics. 
Kitchenham and Pleeger~\cite{kitchenham_software_1996}, in their foreword to a special issue on software quality, eloquently describe the different views of software quality that include a transcendental view (where quality is an ethereal goal we never quite reach), the user view (how the product meets users' needs), the manufacturing view (product quality during production and after delivery), the product view (the product's inherent internal quality), and the value-based view (value to customers or market groups).  These different views also form the basis for an excellent discussion among developers and managers. 

Gilliet et al. also share that quality means different things to different people in different contexts and is impacted by numerous multi-dimensional factors~\cite{gillies_software_2011}. In particular, they note that what quality means in software development is hard to define, and measuring it using only a few quantitative metrics alone can be misleading and may not map to what developers or managers consider to be quality. The five main dimensions we identified in TRUCE correspond closely to the definition of quality provided by Gillies et al. ~\cite{gillies_software_2011}, as
four of the five dimensions in TRUCE were also in their definition (using slightly different terminology):  Only collaboration was not explicitly mentioned in Gillies' definition. Although collaboration support may be considered as an aspect of evolution, our respondents discussed collaboration in a more immediate sense in that work a developer does has to consider the collaborative needs that are pressing (e.g., making quality trade-offs to unblock others). 

Wilson and Hall conducted a study in 1994/95 to compare developer, manager, and software quality practitioner views on software quality from different companies~\cite{wilson1998perceptions}. 
They found ``that management often
appear unaware of their developers’ opinions of software quality''  and that the ``[t]he significant impediments to software quality initiatives are lack of resources and lack of time''.  They also found that managers underestimated the amount of effort developers take to ensure quality. 

\subsection{Where Productivity and Quality Meet}

There has been extensive research that considers both quality and productivity (not surprising given how related they are).  Notably, the performance dimension of the SPACE framework is about the quality of the outcomes achieved, and Sadowski et al. also suggest that quality is an important aspect of developer productivity~\cite{sadowski_software_2019}. Some researchers have considered the impact of interventions on both developer productivity and software quality. For example, Bissi et al. conducted a review of test-driven development approaches on developer productivity and quality ~\cite{bissi2016effects}, Coupe et al. investigated the impact of computer-aided software engineering tools on productivity and quality~\cite{coupe1996empirical}, and Savor et al.~\cite{savor2016} and Vasilescu et al.~\cite{vasilescu2015quality} considered the impact of continuous deployment on productivity, measured by shipped lines of code and number of commits (respectively), and quality, measured by number of failures. 
Other researchers have explored the impact of both productivity and quality perceptions on adoption of improvement processes~\cite{green2005impacts}.  We did not, however, identify other research that has investigated how developers and managers define both quality and productivity, how aligned the two cohorts are in their views, and how they report making trade-offs between them. 

In our research, we found that both developers and managers have different views of both productivity (within and across cohorts) and quality (within cohorts, recall we did not ask survey questions about their perceptions of how the other cohort defined quality), and that over half of over half of our participants report trading quality for productivity.   As Wilson and Hall noted back in 1998, ``Many software quality initiatives fail because they do not take account of the range of views that people have of quality''~\cite{wilson1998perceptions}. They also stressed that ``new approaches to software quality improvement will not work unless software developers believe in them, no matter how enthusiastic managers may be''. We go one step further, and suggest that developers and managers need to discuss their views of developer productivity, software quality, and when/if they should make trade-offs, and why or why not.  

\section{Conclusions}

We found that many developers were not aware how their managers perceived productivity, and many were nervous about having their work measured and evaluated using metrics from telemetry data, especially given the invisible nature of some development activities (e.g., helping others and learning or planning for the future).

The new framing of quality we offer, and the insights on misalignment of what productivity means to developers and managers, suggests a need for developer teams and their managers to regularly reflect and discuss what productivity and quality mean to them. Many companies lately have also been striving to understand and measure productivity and quality; thus, our research calls for attention on how differences in developer and manager views of productivity and quality should be discussed before any metrics or interventions to improve quality and productivity are put into practice.  Quality criteria in any domain are seldom independent, and making compromises often brings up conflicts~\cite{gillies_software_2011}, but highlighting which aspects of quality should be traded is important to do.  For example, one aspect of quality is timely delivery (and some practitioners advocate this is the most important quality criteria) of software that meets user needs~\cite{kitchenham_software_1996}, but other practitioners recognize that software that is useful will need to evolve, and writing code with evolution in mind is essential~\cite{gillies_software_2011}. Through our study, we offered some feedback (and reinforcement) that SPACE can be used for thinking about developer productivity, and how TRUCE may also be used to have richer conversations about trade-offs that are made in terms of software quality. 

Despite the limitations with using quantitative metrics for measuring both productivity and quality, many companies strive to use metrics to have fast ways to measure changes and drive insights about possible improvements.  Recognizing that metrics will be used despite their known limitations, we suggest using both SPACE and TRUCE for defining metrics. Using both frameworks can help distinguish developer productivity metrics from software quality metrics, and help practitioners identify a more diverse set of metrics that will capture a broader picture of productivity and quality in their companies or projects.

\newlength\myheight
\newlength\mydepth
\settototalheight\myheight{Xygp}
\settodepth\mydepth{Xygp}

\begin{acks}
We thank the survey participants, and 
Jacek Czerwonka, Cassandra Petrachenko and Alessandra Paz Milani 
for feedback on our paper.
Inspired by Courtney Miller and Chanel (Miller @ ICSE 2022), a very special thanks to  Kodabear~\raisebox{-\mydepth}{\includegraphics[height=\baselineskip]{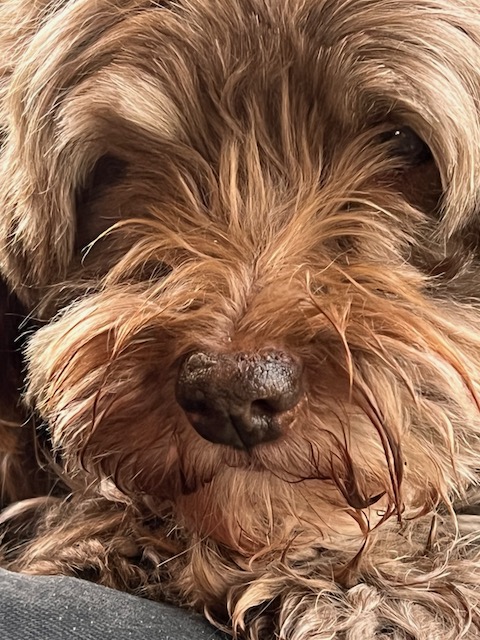}} for keeping the authors company on their many Teams and Zoom calls! \#dogsoficse \#dogsofchase
\end{acks}

\bibliographystyle{ACM-Reference-Format}
\bibliography{main}


\begin{thebibliography}{26}


\ifx \showCODEN    \undefined \def \showCODEN     #1{\unskip}     \fi
\ifx \showDOI      \undefined \def \showDOI       #1{#1}\fi
\ifx \showISBNx    \undefined \def \showISBNx     #1{\unskip}     \fi
\ifx \showISBNxiii \undefined \def \showISBNxiii  #1{\unskip}     \fi
\ifx \showISSN     \undefined \def \showISSN      #1{\unskip}     \fi
\ifx \showLCCN     \undefined \def \showLCCN      #1{\unskip}     \fi
\ifx \shownote     \undefined \def \shownote      #1{#1}          \fi
\ifx \showarticletitle \undefined \def \showarticletitle #1{#1}   \fi
\ifx \showURL      \undefined \def \showURL       {\relax}        \fi
\providecommand\bibfield[2]{#2}
\providecommand\bibinfo[2]{#2}
\providecommand\natexlab[1]{#1}
\providecommand\showeprint[2][]{arXiv:#2}

\bibitem[\protect\citeauthoryear{Beecham, Baddoo, Hall, Robinson, and
  Sharp}{Beecham et~al\mbox{.}}{2008}]%
        {BBHRS08}
\bibfield{author}{\bibinfo{person}{Sarah Beecham}, \bibinfo{person}{Nathan
  Baddoo}, \bibinfo{person}{Tracy Hall}, \bibinfo{person}{Hugh Robinson}, {and}
  \bibinfo{person}{Helen Sharp}.} \bibinfo{year}{2008}\natexlab{}.
\newblock \showarticletitle{Motivation in Software Engineering: A systematic
  literature review}.
\newblock \bibinfo{journal}{\emph{Information and software technology}}
  \bibinfo{volume}{50}, \bibinfo{number}{9-10} (\bibinfo{year}{2008}),
  \bibinfo{pages}{860--878}.
\newblock


\bibitem[\protect\citeauthoryear{Bissi, Neto, and Emer}{Bissi
  et~al\mbox{.}}{2016}]%
        {bissi2016effects}
\bibfield{author}{\bibinfo{person}{Wilson Bissi}, \bibinfo{person}{Adolfo
  Gustavo Serra~Seca Neto}, {and} \bibinfo{person}{Maria Claudia
  Figueiredo~Pereira Emer}.} \bibinfo{year}{2016}\natexlab{}.
\newblock \showarticletitle{The effects of test driven development on internal
  quality, external quality and productivity: A systematic review}.
\newblock \bibinfo{journal}{\emph{Information and Software Technology}}
  \bibinfo{volume}{74} (\bibinfo{year}{2016}), \bibinfo{pages}{45--54}.
\newblock


\bibitem[\protect\citeauthoryear{Coupe and Onodu}{Coupe and Onodu}{1996}]%
        {coupe1996empirical}
\bibfield{author}{\bibinfo{person}{RT Coupe} {and} \bibinfo{person}{NM Onodu}.}
  \bibinfo{year}{1996}\natexlab{}.
\newblock \showarticletitle{An empirical evaluation of the impact of CASE on
  developer productivity and software quality}.
\newblock \bibinfo{journal}{\emph{Journal of Information Technology}}
  \bibinfo{volume}{11}, \bibinfo{number}{2} (\bibinfo{year}{1996}),
  \bibinfo{pages}{173--181}.
\newblock


\bibitem[\protect\citeauthoryear{Forsgren, Storey, Maddila, Zimmermann, Houck,
  and Butler}{Forsgren et~al\mbox{.}}{2021}]%
        {space}
\bibfield{author}{\bibinfo{person}{Nicole Forsgren},
  \bibinfo{person}{Margaret-Anne Storey}, \bibinfo{person}{Chandra Maddila},
  \bibinfo{person}{Thomas Zimmermann}, \bibinfo{person}{Brian Houck}, {and}
  \bibinfo{person}{Jenna Butler}.} \bibinfo{year}{2021}\natexlab{}.
\newblock \showarticletitle{The SPACE of Developer Productivity: There's More
  to It than You Think.}
\newblock \bibinfo{journal}{\emph{Queue}} \bibinfo{volume}{19},
  \bibinfo{number}{1} (\bibinfo{date}{Feb.} \bibinfo{year}{2021}),
  \bibinfo{pages}{20–48}.
\newblock
\showISSN{1542-7730}
\urldef\tempurl%
\url{https://doi.org/10.1145/3454122.3454124}
\showDOI{\tempurl}


\bibitem[\protect\citeauthoryear{Fran{\c{c}}a, Da~Silva, and
  Sharp}{Fran{\c{c}}a et~al\mbox{.}}{2018}]%
        {FDS18}
\bibfield{author}{\bibinfo{person}{C{\'e}sar Fran{\c{c}}a},
  \bibinfo{person}{Fabio~QB Da~Silva}, {and} \bibinfo{person}{Helen Sharp}.}
  \bibinfo{year}{2018}\natexlab{}.
\newblock \showarticletitle{Motivation and satisfaction of software engineers}.
\newblock \bibinfo{journal}{\emph{IEEE Transactions on Software Engineering}}
  (\bibinfo{year}{2018}).
\newblock


\bibitem[\protect\citeauthoryear{Gillies}{Gillies}{2011}]%
        {gillies_software_2011}
\bibfield{author}{\bibinfo{person}{Alan Gillies}.}
  \bibinfo{year}{2011}\natexlab{}.
\newblock \bibinfo{booktitle}{\emph{Software {Quality}: {Theory} and
  {Management}}}.
\newblock \bibinfo{publisher}{Lulu.com}.
\newblock
\showISBNx{978-1-4467-5398-9}
\newblock
\shownote{Google-Books-ID: XTvpAQAAQBAJ}.


\bibitem[\protect\citeauthoryear{Graziotin, Wang, and Abrahamsson}{Graziotin
  et~al\mbox{.}}{2013}]%
        {graziotin2013}
\bibfield{author}{\bibinfo{person}{Daniel Graziotin}, \bibinfo{person}{Xiaofeng
  Wang}, {and} \bibinfo{person}{Pekka Abrahamsson}.}
  \bibinfo{year}{2013}\natexlab{}.
\newblock \showarticletitle{Are happy developers more productive?}. In
  \bibinfo{booktitle}{\emph{International Conference on Product Focused
  Software Process Improvement}}. Springer, \bibinfo{pages}{50--64}.
\newblock


\bibitem[\protect\citeauthoryear{Green, Hevner, and Collins}{Green
  et~al\mbox{.}}{2005}]%
        {green2005impacts}
\bibfield{author}{\bibinfo{person}{Gina~C Green}, \bibinfo{person}{Alan~R
  Hevner}, {and} \bibinfo{person}{Rosann~Webb Collins}.}
  \bibinfo{year}{2005}\natexlab{}.
\newblock \showarticletitle{The impacts of quality and productivity perceptions
  on the use of software process improvement innovations}.
\newblock \bibinfo{journal}{\emph{Information and Software Technology}}
  \bibinfo{volume}{47}, \bibinfo{number}{8} (\bibinfo{year}{2005}),
  \bibinfo{pages}{543--553}.
\newblock


\bibitem[\protect\citeauthoryear{Kitchenham and Pfleeger}{Kitchenham and
  Pfleeger}{1996}]%
        {kitchenham_software_1996}
\bibfield{author}{\bibinfo{person}{B. Kitchenham} {and} \bibinfo{person}{S.L.
  Pfleeger}.} \bibinfo{year}{1996}\natexlab{}.
\newblock \showarticletitle{Software quality: the elusive target [special
  issues section]}.
\newblock \bibinfo{journal}{\emph{IEEE Software}} \bibinfo{volume}{13},
  \bibinfo{number}{1} (\bibinfo{date}{Jan.} \bibinfo{year}{1996}),
  \bibinfo{pages}{12--21}.
\newblock
\showISSN{1937-4194}
\urldef\tempurl%
\url{https://doi.org/10.1109/52.476281}
\showDOI{\tempurl}
\newblock
\shownote{Conference Name: IEEE Software}.


\bibitem[\protect\citeauthoryear{Ko}{Ko}{2019}]%
        {ko2019}
\bibfield{author}{\bibinfo{person}{Amy~J. Ko}.}
  \bibinfo{year}{2019}\natexlab{}.
\newblock \bibinfo{booktitle}{\emph{Why We Should Not Measure Productivity}}.
\newblock \bibinfo{publisher}{Apress}, \bibinfo{address}{Berkeley, CA},
  \bibinfo{pages}{21--26}.
\newblock
\showISBNx{978-1-4842-4221-6}
\urldef\tempurl%
\url{https://doi.org/10.1007/978-1-4842-4221-6_3}
\showDOI{\tempurl}


\bibitem[\protect\citeauthoryear{Meyer, Fritz, Murphy, and Zimmermann}{Meyer
  et~al\mbox{.}}{2014}]%
        {MFMZ14}
\bibfield{author}{\bibinfo{person}{Andr{\'e}~N. Meyer}, \bibinfo{person}{Thomas
  Fritz}, \bibinfo{person}{Gail~C. Murphy}, {and} \bibinfo{person}{Thomas
  Zimmermann}.} \bibinfo{year}{2014}\natexlab{}.
\newblock \showarticletitle{Software Developers' Perceptions of Productivity}.
  In \bibinfo{booktitle}{\emph{Proceedings of the 22th International Symposium
  on Foundations of Software Engineering}}.
\newblock


\bibitem[\protect\citeauthoryear{Meyer, Murphy, Fritz, and Zimmermann}{Meyer
  et~al\mbox{.}}{2019}]%
        {meyer_developers_2019}
\bibfield{author}{\bibinfo{person}{André~N. Meyer}, \bibinfo{person}{Gail~C.
  Murphy}, \bibinfo{person}{Thomas Fritz}, {and} \bibinfo{person}{Thomas
  Zimmermann}.} \bibinfo{year}{2019}\natexlab{}.
\newblock \showarticletitle{Developers' {Diverging} {Perceptions} of
  {Productivity}}.
\newblock In \bibinfo{booktitle}{\emph{Rethinking {Productivity} in {Software}
  {Engineering}}}, \bibfield{editor}{\bibinfo{person}{Caitlin Sadowski} {and}
  \bibinfo{person}{Thomas Zimmermann}} (Eds.). \bibinfo{publisher}{Apress},
  \bibinfo{address}{Berkeley, CA}, \bibinfo{pages}{137--146}.
\newblock
\showISBNx{978-1-4842-4221-6}
\urldef\tempurl%
\url{https://doi.org/10.1007/978-1-4842-4221-6_12}
\showDOI{\tempurl}


\bibitem[\protect\citeauthoryear{Mikkonen}{Mikkonen}{2016}]%
        {mikkonen2016flow}
\bibfield{author}{\bibinfo{person}{Tommi Mikkonen}.}
  \bibinfo{year}{2016}\natexlab{}.
\newblock \showarticletitle{Flow, intrinsic motivation, and developer
  experience in software engineering}.
\newblock \bibinfo{journal}{\emph{Agile Processes in Software Engineering and
  Extreme Programming}} (\bibinfo{year}{2016}), \bibinfo{pages}{104}.
\newblock


\bibitem[\protect\citeauthoryear{Miller, Rodeghero, Storey, Ford, and
  Zimmermann}{Miller et~al\mbox{.}}{2021}]%
        {miller:icse:2021}
\bibfield{author}{\bibinfo{person}{Courtney Miller}, \bibinfo{person}{Paige
  Rodeghero}, \bibinfo{person}{Margaret-Anne Storey}, \bibinfo{person}{Denae
  Ford}, {and} \bibinfo{person}{Thomas Zimmermann}.}
  \bibinfo{year}{2021}\natexlab{}.
\newblock \showarticletitle{"How Was Your Weekend?" Software Development Teams
  Working From Home During COVID-19}. In \bibinfo{booktitle}{\emph{2021
  IEEE/ACM 43rd International Conference on Software Engineering (ICSE)}}.
  \bibinfo{pages}{624--636}.
\newblock


\bibitem[\protect\citeauthoryear{{Murphy-Hill}, {Jaspan}, {Sadowski},
  {Shepherd}, {Phillips}, {Winter}, {Knight}, {Smith}, and
  {Jorde}}{{Murphy-Hill} et~al\mbox{.}}{2019}]%
        {murphyhill2019predicts}
\bibfield{author}{\bibinfo{person}{E. {Murphy-Hill}}, \bibinfo{person}{C.
  {Jaspan}}, \bibinfo{person}{C. {Sadowski}}, \bibinfo{person}{D. {Shepherd}},
  \bibinfo{person}{M. {Phillips}}, \bibinfo{person}{C. {Winter}},
  \bibinfo{person}{A. {Knight}}, \bibinfo{person}{E. {Smith}}, {and}
  \bibinfo{person}{M. {Jorde}}.} \bibinfo{year}{2019}\natexlab{}.
\newblock \showarticletitle{What Predicts Software Developers' Productivity?}
\newblock \bibinfo{journal}{\emph{IEEE Transactions on Software Engineering}}
  (\bibinfo{year}{2019}).
\newblock
\urldef\tempurl%
\url{https://doi.org/10.1109/TSE.2019.2900308}
\showDOI{\tempurl}


\bibitem[\protect\citeauthoryear{Nistala, Nori, and Reddy}{Nistala
  et~al\mbox{.}}{2019}]%
        {nistala_software_2019}
\bibfield{author}{\bibinfo{person}{Padmalata Nistala},
  \bibinfo{person}{Kesav~Vithal Nori}, {and} \bibinfo{person}{Raghu Reddy}.}
  \bibinfo{year}{2019}\natexlab{}.
\newblock \showarticletitle{Software {Quality} {Models}: {A} {Systematic}
  {Mapping} {Study}}. In \bibinfo{booktitle}{\emph{2019 {IEEE}/{ACM}
  {International} {Conference} on {Software} and {System} {Processes}
  ({ICSSP})}}. \bibinfo{pages}{125--134}.
\newblock


\bibitem[\protect\citeauthoryear{Pradhan, Nanniyur, Melanahalli, Palla, and
  Chulani}{Pradhan et~al\mbox{.}}{2019}]%
        {pradhan_quality_2019}
\bibfield{author}{\bibinfo{person}{Satya Pradhan}, \bibinfo{person}{Venky
  Nanniyur}, \bibinfo{person}{Paddu Melanahalli}, \bibinfo{person}{Munir
  Palla}, {and} \bibinfo{person}{Sunita Chulani}.}
  \bibinfo{year}{2019}\natexlab{}.
\newblock \showarticletitle{Quality {Metrics} for {Hybrid} {Software}
  {Development} {Organizations} – {A} {Case} {Study}}. In
  \bibinfo{booktitle}{\emph{2019 {IEEE} 19th {International} {Conference} on
  {Software} {Quality}, {Reliability} and {Security} {Companion} ({QRS}-{C})}}.
  \bibinfo{pages}{505--506}.
\newblock
\urldef\tempurl%
\url{https://doi.org/10.1109/QRS-C.2019.00097}
\showDOI{\tempurl}


\bibitem[\protect\citeauthoryear{Sadowski, Storey, and Feldt}{Sadowski
  et~al\mbox{.}}{2019}]%
        {sadowski_software_2019}
\bibfield{author}{\bibinfo{person}{Caitlin Sadowski},
  \bibinfo{person}{Margaret-Anne Storey}, {and} \bibinfo{person}{Robert
  Feldt}.} \bibinfo{year}{2019}\natexlab{}.
\newblock \showarticletitle{A {Software} {Development} {Productivity}
  {Framework}}.
\newblock In \bibinfo{booktitle}{\emph{Rethinking {Productivity} in {Software}
  {Engineering}}}, \bibfield{editor}{\bibinfo{person}{Caitlin Sadowski} {and}
  \bibinfo{person}{Thomas Zimmermann}} (Eds.). \bibinfo{publisher}{Apress},
  \bibinfo{address}{Berkeley, CA}, \bibinfo{pages}{39--47}.
\newblock
\showISBNx{978-1-4842-4221-6}
\urldef\tempurl%
\url{https://doi.org/10.1007/978-1-4842-4221-6_5}
\showDOI{\tempurl}


\bibitem[\protect\citeauthoryear{Savor, Douglas, Gentili, Williams, Beck, and
  Stumm}{Savor et~al\mbox{.}}{2016}]%
        {savor2016}
\bibfield{author}{\bibinfo{person}{Tony Savor}, \bibinfo{person}{Mitchell
  Douglas}, \bibinfo{person}{Michael Gentili}, \bibinfo{person}{Laurie
  Williams}, \bibinfo{person}{Kent Beck}, {and} \bibinfo{person}{Michael
  Stumm}.} \bibinfo{year}{2016}\natexlab{}.
\newblock \showarticletitle{Continuous Deployment at Facebook and OANDA}. In
  \bibinfo{booktitle}{\emph{2016 IEEE/ACM 38th International Conference on
  Software Engineering Companion (ICSE-C)}}. \bibinfo{pages}{21--30}.
\newblock


\bibitem[\protect\citeauthoryear{Sharp, Baddoo, Beecham, Hall, and
  Robinson}{Sharp et~al\mbox{.}}{2009}]%
        {SBBHR09}
\bibfield{author}{\bibinfo{person}{Helen Sharp}, \bibinfo{person}{Nathan
  Baddoo}, \bibinfo{person}{Sarah Beecham}, \bibinfo{person}{Tracy Hall}, {and}
  \bibinfo{person}{Hugh Robinson}.} \bibinfo{year}{2009}\natexlab{}.
\newblock \showarticletitle{Models of motivation in software engineering}.
\newblock \bibinfo{journal}{\emph{Information and software technology}}
  \bibinfo{volume}{51}, \bibinfo{number}{1} (\bibinfo{year}{2009}),
  \bibinfo{pages}{219--233}.
\newblock


\bibitem[\protect\citeauthoryear{Storey, Houck, and Zimmermann}{Storey
  et~al\mbox{.}}{2021a}]%
        {supplemental-materials-alignment}
\bibfield{author}{\bibinfo{person}{Margaret-Anne Storey},
  \bibinfo{person}{Brian Houck}, {and} \bibinfo{person}{Thomas Zimmermann}.}
  \bibinfo{year}{2021}\natexlab{a}.
\newblock \bibinfo{booktitle}{\emph{Appendix to How Developers and Managers
  Define and Trade Off Productivity and Quality}}.
\newblock \bibinfo{type}{{T}echnical {R}eport} MSR-TR-2021-26.
  \bibinfo{institution}{Microsoft Research}.
\newblock
\urldef\tempurl%
\url{https://www.microsoft.com/en-us/research/publication/appendix-to-productivity-quality-alignment}
\showURL{%
\tempurl}


\bibitem[\protect\citeauthoryear{Storey, Zimmermann, Bird, Czerwonka, Murphy,
  and Kalliamvakou}{Storey et~al\mbox{.}}{2021b}]%
        {storey_towards_2019}
\bibfield{author}{\bibinfo{person}{Margaret-Anne Storey},
  \bibinfo{person}{Thomas Zimmermann}, \bibinfo{person}{Christian Bird},
  \bibinfo{person}{Jacek Czerwonka}, \bibinfo{person}{Brendan Murphy}, {and}
  \bibinfo{person}{Eirini Kalliamvakou}.} \bibinfo{year}{2021}\natexlab{b}.
\newblock \showarticletitle{Towards a {Theory} of {Software} {Developer} {Job}
  {Satisfaction} and {Perceived} {Productivity}}.
\newblock \bibinfo{journal}{\emph{IEEE Transactions on Software Engineering}}
  \bibinfo{volume}{47}, \bibinfo{number}{10} (\bibinfo{year}{2021}),
  \bibinfo{pages}{2125--2142}.
\newblock


\bibitem[\protect\citeauthoryear{Van~Solingen, Basili, Caldiera, and
  Rombach}{Van~Solingen et~al\mbox{.}}{2002}]%
        {basili}
\bibfield{author}{\bibinfo{person}{Rini Van~Solingen}, \bibinfo{person}{Vic
  Basili}, \bibinfo{person}{Gianluigi Caldiera}, {and}
  \bibinfo{person}{H~Dieter Rombach}.} \bibinfo{year}{2002}\natexlab{}.
\newblock \showarticletitle{Goal question metric (gqm) approach}.
\newblock \bibinfo{journal}{\emph{Encyclopedia of software engineering}}
  (\bibinfo{year}{2002}).
\newblock


\bibitem[\protect\citeauthoryear{Vasilescu, Yu, Wang, Devanbu, and
  Filkov}{Vasilescu et~al\mbox{.}}{2015}]%
        {vasilescu2015quality}
\bibfield{author}{\bibinfo{person}{Bogdan Vasilescu}, \bibinfo{person}{Yue Yu},
  \bibinfo{person}{Huaimin Wang}, \bibinfo{person}{Premkumar Devanbu}, {and}
  \bibinfo{person}{Vladimir Filkov}.} \bibinfo{year}{2015}\natexlab{}.
\newblock \showarticletitle{Quality and productivity outcomes relating to
  continuous integration in GitHub}. In \bibinfo{booktitle}{\emph{Proceedings
  of the 2015 10th Joint Meeting on Foundations of Software Engineering}}.
  \bibinfo{pages}{805--816}.
\newblock


\bibitem[\protect\citeauthoryear{Wagner and Ruhe}{Wagner and Ruhe}{2018}]%
        {Wagner2018}
\bibfield{author}{\bibinfo{person}{Stefan Wagner} {and}
  \bibinfo{person}{Melanie Ruhe}.} \bibinfo{year}{2018}\natexlab{}.
\newblock \showarticletitle{A systematic review of productivity factors in
  software development}.
\newblock \bibinfo{journal}{\emph{arXiv preprint arXiv:1801.06475}}
  (\bibinfo{year}{2018}).
\newblock


\bibitem[\protect\citeauthoryear{Wilson and Hall}{Wilson and Hall}{1998}]%
        {wilson1998perceptions}
\bibfield{author}{\bibinfo{person}{David~N Wilson} {and} \bibinfo{person}{Tracy
  Hall}.} \bibinfo{year}{1998}\natexlab{}.
\newblock \showarticletitle{Perceptions of software quality: a pilot study}.
\newblock \bibinfo{journal}{\emph{Software quality journal}}
  \bibinfo{volume}{7}, \bibinfo{number}{1} (\bibinfo{year}{1998}),
  \bibinfo{pages}{67--75}.
\newblock


\end{thebibliography}

\end{document}